%% file: 0rtt_satcom.tex
\documentclass[conference]{IEEEtran}
\IEEEoverridecommandlockouts
\usepackage{cite}
\usepackage{amsmath,amssymb,amsfonts}
\usepackage{algorithmic}
\usepackage{graphicx}
\usepackage{textcomp}
\usepackage{xcolor}
\usepackage{tikz}
\usepackage{pgfplots}
\usepackage{todonotes}
\usepackage{tabularx}

\def\BibTeX{{\rm B\kern-.05em{\sc i\kern-.025em b}\kern-.08em
    T\kern-.1667em\lower.7ex\hbox{E}\kern-.125emX}}
\tikzset{every picture/.style={line width=0.75pt}}
\tikzset{every picture/.style={line width=0.75pt}}
\tikzset{cross/.style={cross out, draw=black, fill=none, minimum size=2*(#1-\pgflinewidth), inner sep=0pt, outer sep=0pt}, cross/.default={2pt}}
\usetikzlibrary{positioning, shapes, trees, automata, calc, fit, automata, shapes.misc, arrows, patterns}
\begin{document}
\title{Evaluating $BDP\_FRAME$ extension for QUIC}

\author{
\IEEEauthorblockN{Nicolas Kuhn}
\IEEEauthorblockA{\textit{CNES}}
\and
\IEEEauthorblockN{Francklin Simo}
\IEEEauthorblockA{\textit{VIVERIS TECHNOLOGIES}}
\and
\IEEEauthorblockN{David Pradas}
\IEEEauthorblockA{\textit{VIVERIS TECHNOLOGIES}}
\and
\IEEEauthorblockN{Emile Stephan}
\IEEEauthorblockA{\textit{ORANGE LABS}}
}

\maketitle

\begin{abstract}

The first version of QUIC has recently been standardized by the IETF. The
	framework of QUIC enables the proposition, negociation and exploitation
	of extensions to adapt some of its mechanisms. As one example, the
	DATAGRAM extension enables the unreliable transmission of data.  

The $BDP\_FRAME$ extension is a method that can improve traffic delivery by
	allowing a QUIC connection to remember the knowledge of path
	characteristics and exploit them when resuming a session.

This technical report presents the rationale behind fast convergence in SATCOM
	systems and evaluate the $BDP\_FRAME$ extension in emulated and live
	environments.

\end{abstract}


\input{introduction}

\newpage
\input{parameters}

\newpage
\input{need_convergence}

\newpage
\input{0rttbdp_single_flow}
\newpage
\input{0rttbdp_congestion}

\newpage
\input{conclusion}
\input{ackno}

\bibliographystyle{IEEEtran}
\bibliography{reference}

\end{document}

%% file: introduction.tex
\section{Introduction}
\label{sec:intro} 

The protocols deployed in the extremities can hardly be relevant for each of
the links available on the Internet, due to their diversity ranging from ``very
high speed low latency'' from data centers to ``high throughput high latency''
from satellite systems. 

Systems exploiting satellites in geostationary orbit see their throughput
increased to provide offers comparable to those of terrestrial systems. This
increase in throughput, combined with the intrinsic latency of these systems,
impacts congestion controls such as TCP.  In order to make full use of the
available capacity, these systems split the TCP connections into sub-segments
to use suitable congestion control on the satellite segment~\cite{RFC3135}. 

These solutions can be applied at different levels of the protocol stack and
are generally at the level of the transport layer in SATellite COMmunication
(SATCOM), and in in particular by adapting the Transmission Control Protocol
(TCP).  With a TCP PEP, the packet losses are distributed over three
sub-segments and the control of congestion can be adapted on the satellite
link. This can result in a reduction by half the loading time of a web page. 

The end-to-end deployment of QUIC challenges these adaptations. Indeed, with
QUIC (RFC8999~\cite{RFC8999}, RFC9000~\cite{RFC9000} and RFC9001~\cite{RFC9001}), the
functionalities previously distributed between Hypertext Transfer Protocol
(HTTP)1/1.1/2, Transport Layer Security (TLS) and TCP are shared between
HTTP3, QUIC and UDP. As the UDP protocol is not in connected mode, it is not
possible to cut out end-to-end communication as with TCP. In addition, QUIC was
measured as taking a not insignificant part of the volume transmitted in a
broadband access, in particular because of the actors implementing it, which
are Google or Facebook. It is therefore necessary understand the performance of
the QUIC protocol in a deployment context that has not been, a priori,
considered in its design.

The authors of~\cite{9268814} identify that the main challenges ahead of QUIC
in SATCOM systems are: (1) tolerance to packet loss, (2) adapted buffer sizes at
the end points, (3) adequate congestion control and (4) quickly exploiting the
available capacity. As the server does not have a priori knowledge of the
characteristics of the underlying system, the convergence of congestion control
can introduce a significant delay before the flow of the communication is
operated. An extension do QUIC currently under discussion at the
IETF\cite{0rttdraft} is to record the RTT and congestion window parameters
during a first connection and send them to the client in a $BDP\_FRAME$.
When the client wishes to reconnect to the server, the client returns this
$BDP\_FRAME$ and the parameters of the previous connection can be used.

This technical report presents the rationale behind fast convergence in SATCOM
systems and evaluate the 0rtt-bdp extension in emulated and live environments.

%% file: parameters.tex
\section{Configurations}
\label{configurations} 

This section presents the different configurations that are considered in this
technical report. 

\begin{figure}[h]
        \centering
        \includegraphics[width =0.8\linewidth]{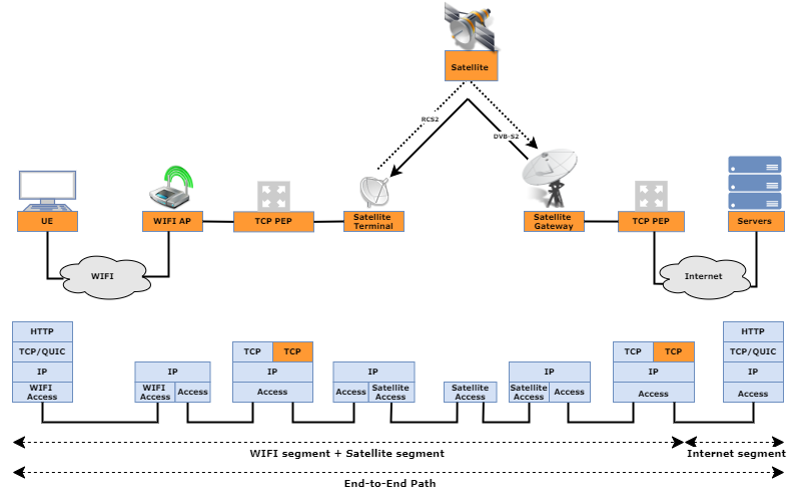}
        \caption{End-to-end SATCOM broadband access - with TCP-PEP}
        \label{fig:protocol_stack_pep}
\end{figure}

\begin{figure}[h]
        \centering
        \includegraphics[width =0.8\linewidth]{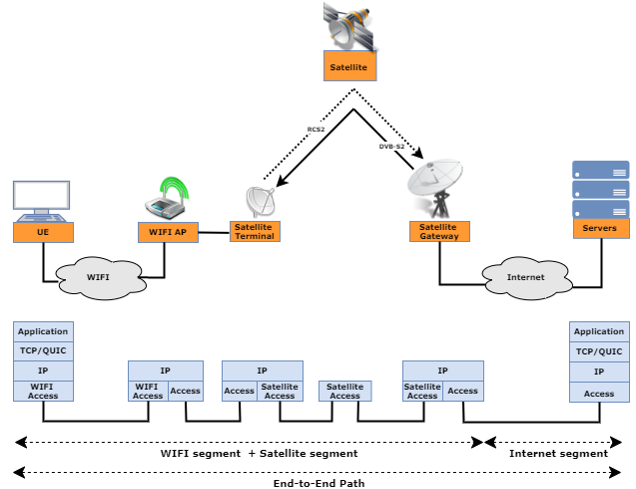}
        \caption{End-to-end SATCOM broadband access - with QUIC}
        \label{fig:protocol_stack_no_pep}
\end{figure}

The generic SATCOM system for a broadband access is shown in
Figure~\ref{fig:protocol_stack_pep}. When QUIC is exploited at the end points, the TCP-PEP component disapears 
as shown in Figure~\ref{fig:protocol_stack_no_pep}. 

\begin{figure}[h]
        \centering
        \includegraphics[width =0.7\linewidth]{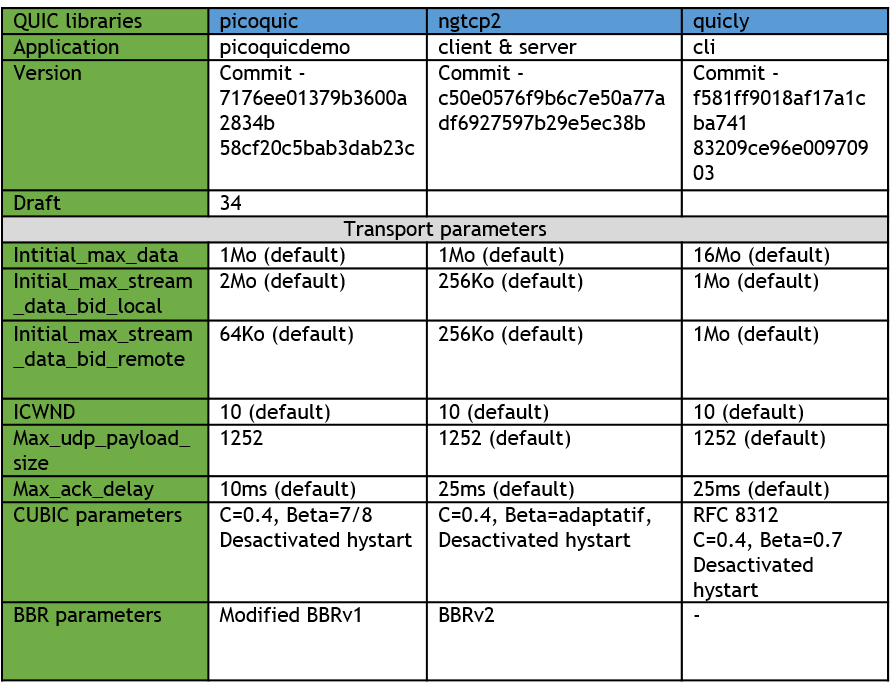}
        \caption{QUIC implementations and configurations}
        \label{fig:quic_stack}
\end{figure}

\begin{figure}[h]
        \centering
        \includegraphics[width =0.7\linewidth]{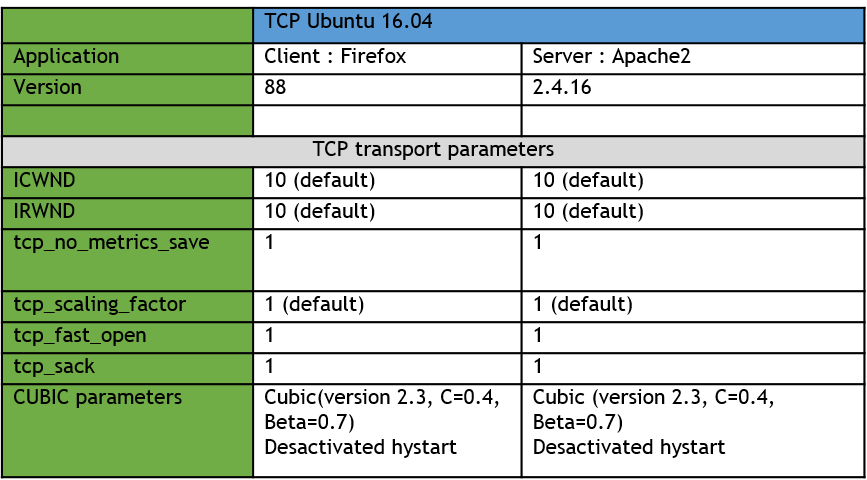}
        \caption{TCP end points configurations}
        \label{fig:tcp_srv_clt}
\end{figure}

\begin{figure}[h]
        \centering
        \includegraphics[width =0.7\linewidth]{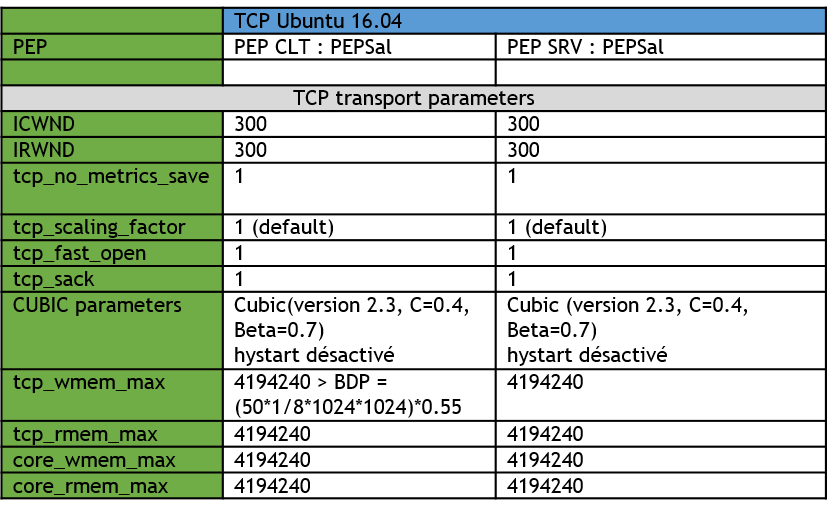}
        \caption{PEP configurations}
        \label{fig:tcp_pep_srv_clt}
\end{figure}

The configuration of the considered QUIC implementations are shown in
Figure~\ref{fig:quic_stack}, the configuration of the end points TCP in
Figure~\ref{fig:tcp_srv_clt} and the configuration of the PEP in
Figure~\ref{fig:tcp_pep_srv_clt}. The PEP exploits PEPSal.

Unless specify otherwise, the bottleneck bandwidth on the forward link of the
SATCOM system is set to $50$\,Mbps and the return link to $10$\,Mbps.

%% file: need_convergence.tex
\section{Rationale behind congestion control convergence}
\label{sec:need_convergence}

This section provides some emulation results to illustrate the rationale
behind better congestion control convergence and the relation between achieved
throughput and transmitted file size. 

The tests presented in this section exploits a two peers network. More details
on the set up and the exploited platform can be found in
GITHUB~\footnote{https://github.com/NicoKos/openbach-example-simple}.

\begin{figure}[h]
        \centering
        \includegraphics[width =0.7\linewidth]{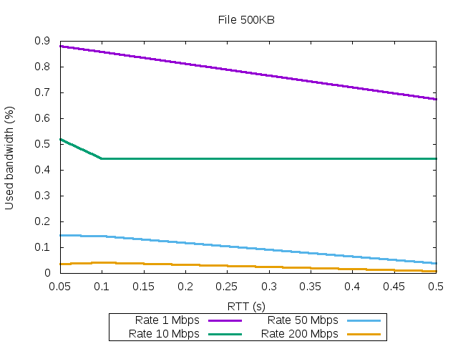}
        \caption{500KB file size - used bandwidth as a function of the RTT for various bottleneck data rates}
	\label{fig:500kB_need_convergence}
\end{figure}

\begin{figure}[h]
        \centering
        \includegraphics[width =0.7\linewidth]{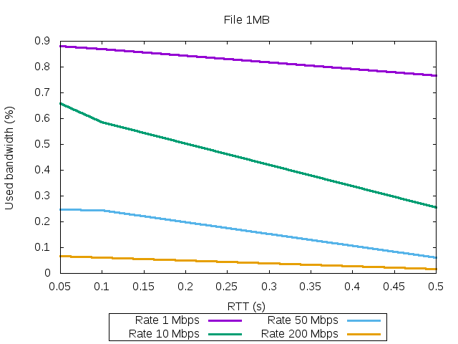}
        \caption{1MB file size - used bandwidth as a function of the RTT for various bottleneck data rates}
	\label{fig:1MB_need_convergence}
\end{figure}

\begin{figure}[h]
        \centering
        \includegraphics[width =0.7\linewidth]{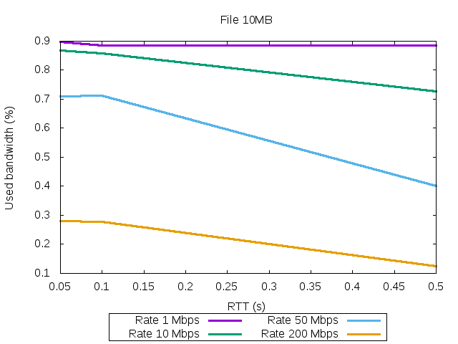}
        \caption{10MB file size - used bandwidth as a function of the RTT for various bottleneck data rates}
	\label{fig:10mb_need_convergence} 
\end{figure}

\begin{figure}[h]
        \centering
        \includegraphics[width =0.7\linewidth]{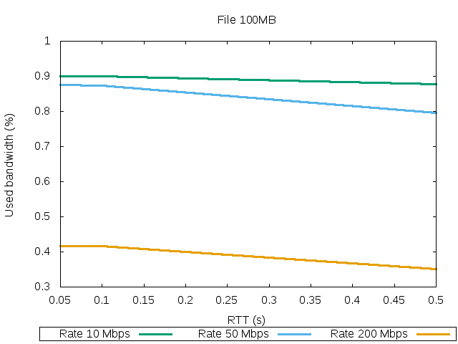}
        \caption{100MB file size - used bandwidth as a function of the RTT for various bottleneck data rates}
	\label{fig:100mb_need_convergence}
\end{figure}

QUIC implementation in picoquic using BBR congestion control is exploited to
transfer 500 KB, 1 MB, 10 MB and 100 MB files over RTT $\in$ [100 ms; 500 ms] and
bottlenecks of 1 Mbps (forward) / 100 kbps (return), 10 Mbps (forward) / 2 Mbps
(return), 50 Mbps (forward) / 25 Mbps (return) and 200 Mbps (forward) / 100
Mbps (return). Figures~\ref{fig:500kB_need_convergence},
~\ref{fig:1MB_need_convergence}, ~\ref{fig:10mb_need_convergence} and
~\ref{fig:100mb_need_convergence} represent the used bandwidth for various
forward link bottleneck rate as a function of the RTT. The used bandwidth is
computed as the ratio between the experienced goodput (ratio between the file
size and the time it required to download) and the bottleneck data rate.

The main conclusions are the following: 
\begin{itemize}
	\item with a 10 MB file and a data rate of 1 Mbps, the bottleneck is
		used for all RTT;
	\item for shorter files (e.g. 1 MB), increasing the RTT severely
		impacts link utilization;
	\item when the data rate is high (e.g. 250 Mbps), even a 100 MB
		transfer does not utilize the available capacity;
	\item increasing the file size increases the link utilization.
\end{itemize}

Moreover, the results presented in this section show that there are many cases
where the available capacity is not exploited. This is the case for large RTT
but also the case for small RTT. The convergence of the congestion control has
an impact on the transfer time and in general, the simple estimation of the
transfer time using the file size and the data rate is wrong. Indeed, this
estimation considers instant convergence of the congestion control.

%% file: 0rttbdp_single_flow.tex
\section{Performance improvements of $BDP\_FRAME$}
\label{sec:perfo}

\subsection{$BDP\_FRAME$ extension in a nutshell}

\begin{figure}[h]
        \centering
        \includegraphics[width =\linewidth]{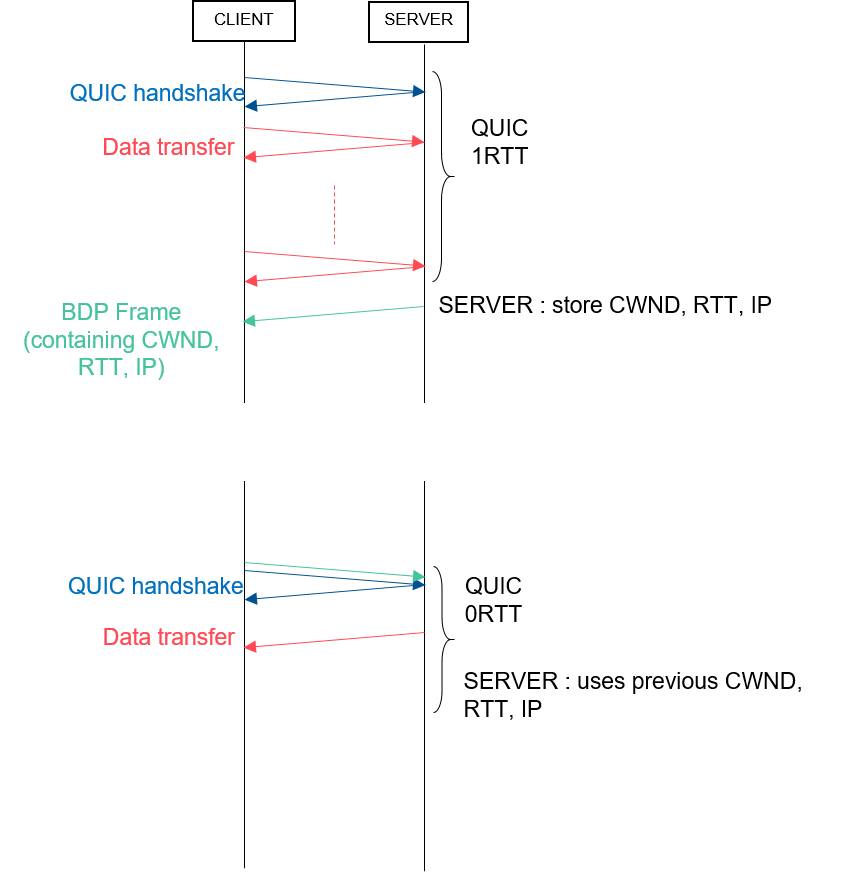}
        \caption{Illustration of the $BDP\_FRAME$ exploitation}
        \label{fig:0rtt_bdp_presentation}
\end{figure}

\cite{0rttdraft} presents three methods to exploit the transport parameters of
a previous session when resuming a session: (1) local storage, where the server
stores the parameters without negociations with the client, (2) NEW\_TOKEN,
where a token that the client can not read is exploited and (3) $BDP\_FRAME$
extension, where a token that the client can read is exploited. The core idea
of the $BDP\_FRAME$ extension is illustrated in
Figure~\ref{fig:0rtt_bdp_presentation} and resides in the following : 

\begin{itemize}
	\item during a previous session, the RTT, the congestion window and the
		IP of the peers are registered by the server and stored in a
		$BDP\_FRAME$;
	\item during this same previous session or at the end of it, the
		$BDP\_FRAME$ is sent to the client;
	\item when resuming a session to the same server, the client sends the
		$BDP\_FRAME$ to the server;
	\item the server can exploit the parameters contained in the
		$BDP\_FRAME$ to adapt the congestion control parameters.
\end{itemize}

More details on the $BDP\_FRAME$ and the ``0RTTBDP'' activity in QUIC can be
found in~\cite{0rttdraft}. The local storage option, where the server stores
the parameters when resuming a session is implemented in picoquic, along with a
safety check that the parameters have not change since previous session. The
$BDP\_FRAME$ option is also available in picoquic.  

\subsection{Emulated performance of $BDP\_FRAME$}

\begin{figure}[h]
        \centering
        \includegraphics[width =0.7\linewidth]{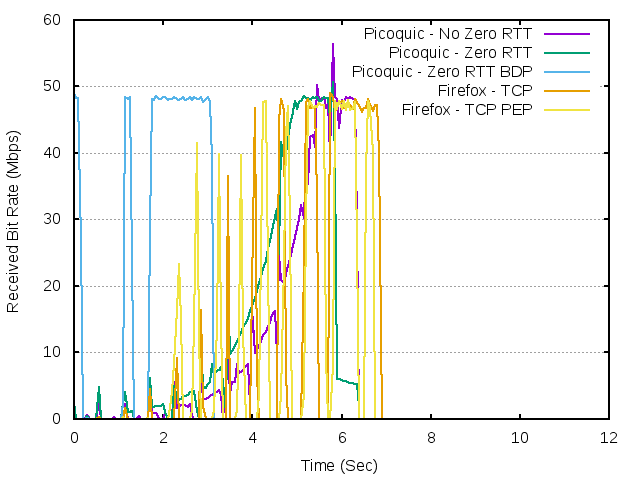}
        \caption{Received data as a function of time with QUIC or TCP, with or without 0RTT}
        \label{fig:illustration_bdp_frame}
\end{figure}

Figure~\ref{fig:illustration_bdp_frame} presents the received data rate as 
a function of time for QUIC, QUIC with 0RTT, QUIC with the $BDP\_FRAME$ extension,
TCP with a PEP and TCP without a PEP. This illustrates that the 0RTT enables a faster
connection establishment than QUIC without 0RTT but also illustrates how the 
$BDP\_FRAME$ furthers helps in quickly exploiting the available capacity. 

\begin{figure}[h]
        \centering
        \includegraphics[width =0.7\linewidth]{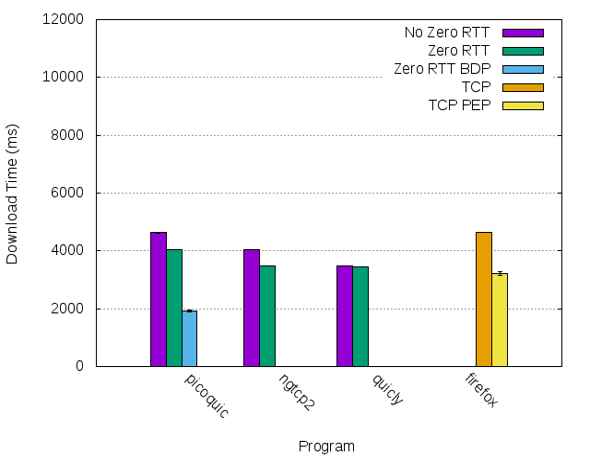}
        \caption{500KB transfer time in SATCOM context with QUIC or TCP, with or without 0RTT}
        \label{fig:500k_transfer_download_time}
\end{figure}

Figure~\ref{fig:500k_transfer_download_time} shows the transfer time of
$500$\,KB with three different implementations of QUIC with and without 0RTT,
with TCP and with TCP-PEP.  

A good news for SATCOM systems is that the three QUIC implementations provide
results close to those of TCP. There are even cases (quicly without 0RTT or
quicly with 0RTT or ngtcp2) where QUIC performs as good as TCP-PEP
configuration. 

The download time is reduced by 13 \% with the usage of the 0RTT with ngtcp2 and
the reduction is neglectible with quicly. Despite the important RTT of SATCOM
systems, using the 0RTT does not seem to contribute much to the reduction of
the transfer time of short files.

With picoquic implementation of QUIC, the exploitation of the 0RTT reduces the
transfer time by 14 \% and by 58 \% with the $BDP\_FRAME$ extension.

%% file: 0rttbdp_congestion.tex
\section{0RTTBDP and variable network conditions}
\label{sec:variable_network}

Section~\ref{sec:perfo} measured that the $BDP\_FRAME$ extension can result in
very important reduction of file transfer time. However, the approach may be
considered as aggressive and it is necessary to assess its performance in
presence of competitive flows.

\subsection{Performance of $BDP\_FRAME$ in congested environments}

\begin{figure}[h]
        \centering
        \includegraphics[width =0.7\linewidth]{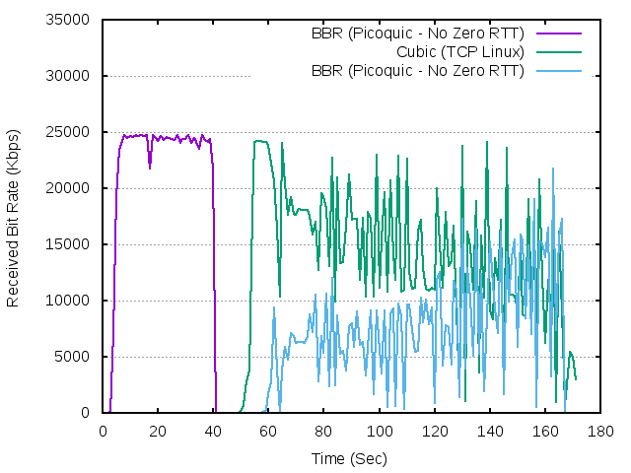}
        \caption{Resume session without 0RTT}
        \label{fig:congestion_no_0rtt}
\end{figure}

\begin{figure}[h]
        \centering
        \includegraphics[width =0.7\linewidth]{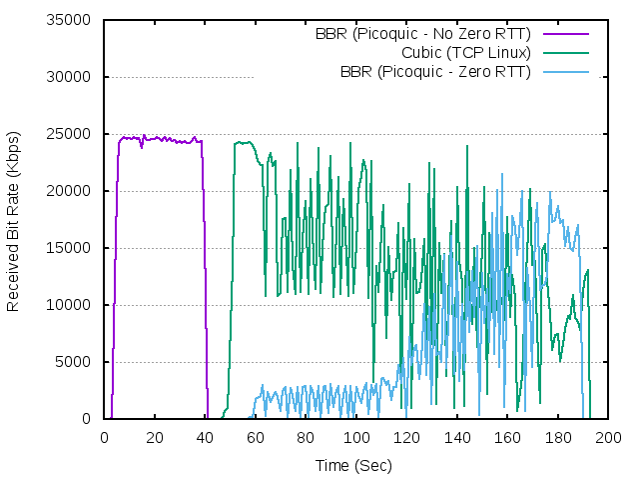}
        \caption{Resume session with 0RTT}
        \label{fig:congestion_0rtt}
\end{figure}

\begin{figure}[h]
        \centering
        \includegraphics[width =0.7\linewidth]{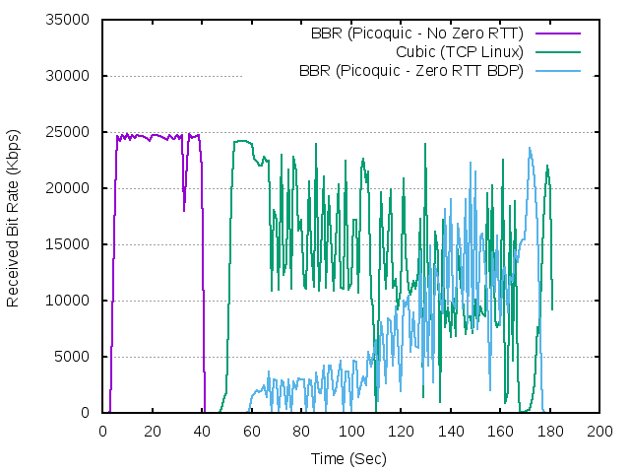}
        \caption{Resume session with $BDP\_FRAME$}
        \label{fig:congestion_0rtt_bdp}
\end{figure}

The scenario that is considered in this section is the following.  At $t=0$ seconds, a first
QUIC connection transmits 100 MB between two peers (purple curve). Then, at $t=60$ seconds, the QUIC connection resumes the connection (blue curve). 
However, in the meantime, at $t=45$ seconds, a TCP flow has grabbed the
available capacity (green curve). This scenario mesures the received bit rate as a function
of time in Figure~\ref{fig:congestion_no_0rtt} (without 0RTT),
Figure~\ref{fig:congestion_0rtt} (with 0RTT) and
Figure~\ref{fig:congestion_0rtt_bdp} (with $BDP\_FRAME$). As a reminder, the
0RTT variant also exploits the characteristics of the previous session and both
0RTT and $BDP\_FRAME$ variants introduce safety checks as those proposed
in~\cite{0rttdraft}.

Contrary to what could have been expected, because the exploitation of previous
session parameters comes with safety nets, the resumed session is less agressiv
than a standard slow-start approach.

\subsection{Performance of $BDP\_FRAME$ in real life}

\begin{figure}[h!]
        \centering
        \includegraphics[width =\linewidth]{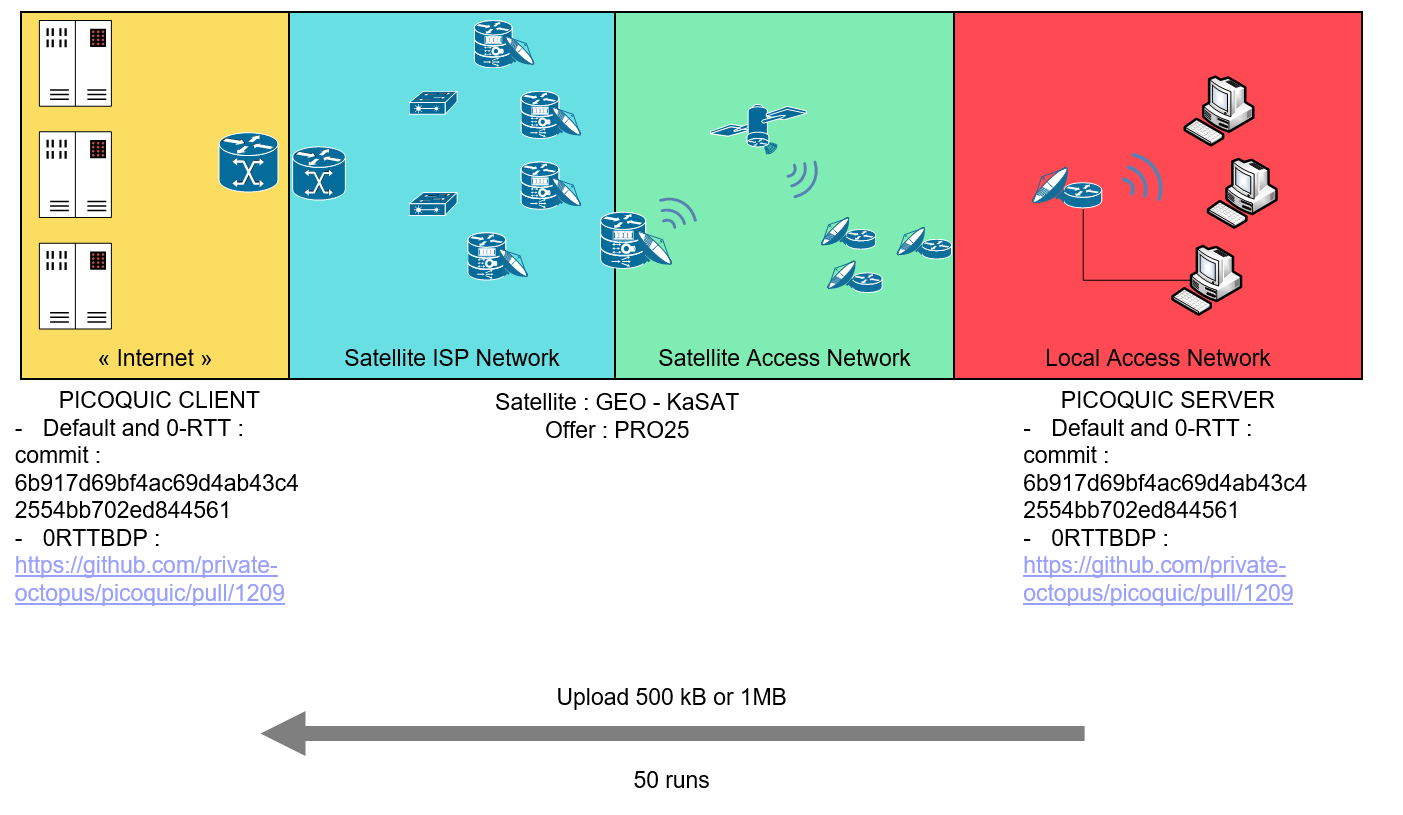}
        \caption{Real satellite acess : architecture}
        \label{fig:real_sat_archi}
\end{figure}

To further assess the performance of the $BDP\_FRAME$ and its deployability, we
have tested it using a real satellite broadband access. The exploitated
architecture is shown in Figure~\ref{fig:real_sat_archi}.

The picoquic server is hosted in private owned servers on the Internet. We
exploited the KaSat satellite and upload 500 KB or 1 MB files, 50 times each,
and use different picoquic clients.

The results of the experiments are shown in Table~\ref{tab:real_sat_results}.
Using the 0RTT as opposed to the 1RTT can result in up to 33 \% reduction in transfer time for 500 KB. With the exploitation of the $BDP\_FRAME$, the transfer time is reduced by up to 67 \%.

\begin{center}
	\begin{table}[h!]
	\centering
	\caption{Real satellite acess : Download Time (DT) of 500 kB and 1 MB}
	\label{tab:real_sat_results}
	\begin{tabular}{ c|c|c|c|c|c|c } 
		& \multicolumn{2}{c|}{Without 0RTT} & \multicolumn{2}{c|}{With 0RTT} & \multicolumn{2}{c}{With $BDP\_FRAME$} \\ \cline{2-7}
	File size (MB) & 0.5 & 1 & 0.5 & 1 & 0.5 & 1 \\ \cline{1-7}
		Min DT (s) & 3.12 & 3.87 & 2.43 & 3.19 & 1.87 & 2.78 \\
		Avg DT (s) & 11.34 & 17.15 & 7.59 & 10.24 & 4.24 & 6.47 \\
		Max DT (s) & 47.82 & 61.43 & 33.69 & 33.92 & 15.55 & 23.88 \\
	\end{tabular}
	\end{table}
\end{center}


%% file: conclusion.tex
\section{Conclusion}
\label{sec:conclu}

This technical report illustrates the need for considering congestion control convergence when determining the transfer time of short objects. Just dividing the file size by the bottleneck data rate may provide wrong assumptions on the actual transfer time.

To improve the convergence of the congestion control in SATCOM environments,
this technical report measures the performance of the $BDP\_FRAME$ extension
for QUIC that exploits the previously measured path characteristics when
resuming a session. With picoquic implementation of QUIC, the exploitation of
the 0RTT reduces the transfer time by 14 \% and by 58 \% with the $BDP\_FRAME$
extension.

The approach may be considered as aggressive and it is necessary to assess its
performance in presence of competitive flows. The evaluations illustrate that
exploiting the previous parameters with a safety check can be less agressive
than a standard slow-start mechanism.

%% file: ackno.tex
\section*{Acknowlegments}

The authors would like to thank Olivier Martinez for his help in setting up the
architecture for the real satellite experiments. The authors thank Christian
Huitema for his fruitfull contribution to the 0RTTBDP and the implementation of
the safety guidelines in picoquic. The authors would like to thank 
Gorry Fairhurst and Tom Jones for their significant discussion on
0RTTBDP.